\documentclass[a4paper]{jpconf}
\usepackage{graphicx}
\begin{document}
\title{Shell model description of Ge isotopes}
\author{J. G. Hirsch and P. C. Srivastava}
\address{Instituto de Ciencias Nucleares, Universidad Nacional Aut\'onoma de
M\'exico, 04510 M\'exico, D.F., M\'exico}
\date{\today}

\ead{hirsch@nucleares.unam.mx}

\begin{abstract}
A shell model study of the low energy region of the spectra in  Ge isotopes for $38\leq N\leq 50$ is presented, analyzing the excitation energies, quadrupole moments, $B(E2)$ values and occupation numbers. The theoretical results have been compared with the available experimental data. The shell model calculations have been performed  employing three different effective interactions and valence spaces.
We have used two effective shell model interactions, JUN45 and jj44b,  for the valence space  $f_{5/2} \, p \,g_{9/2}$ without truncation.
To include the proton subshell $f_{7/2}$ in valence space we have employed the $fpg$ effective interaction due to Sorlin {\it et al.}, with  $^{48}$Ca as a core and a truncation in the number of excited particles.
\end{abstract}
%============================================================================
\section{Introduction}

Ge isotopes exhibit a complex structure in their low energy spectra. The first excited $0^+_2$ state shows an irregular behavior as function of the neutron number, as can be seen in Fig. 1.  From  $^{70}$Ge to  $^{72}$Ge the $0_{2}^+$ level drops in energy, rising from  $^{74}$Ge onwards. In the semimagic isotope $^{72}$Ge, which has $N=40$, the $0_{2}^+$ is below than the $2_{1}^+$. It has been interpreted in terms of configuration mixing of regular and intruder structures, leading to a shape phase transition from oblate to spherical to prolate deformed \cite{Lec80,Toh02,rodal05,rodal10}. 
 
 \begin{figure}[h!]
\begin{center}
\includegraphics[scale=0.455]{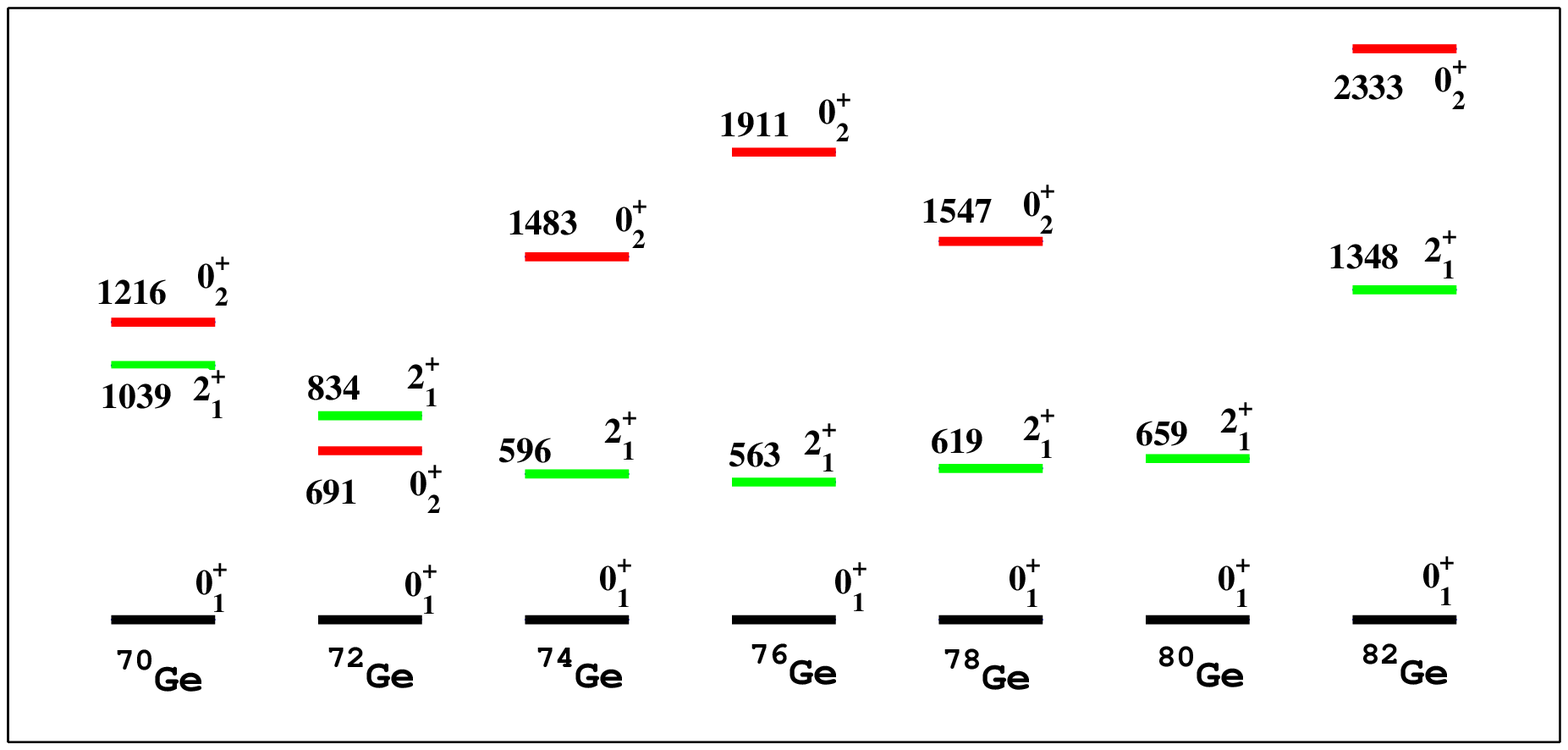}
\caption{(Color online)Low-lying systematics of germanium isotopes.}
\end{center}
\end{figure} 
 
 The double beta decay of $^{76}$Ge plays an upstanding role in the quest to determine the Majorana mass of the neutrino \cite{Ell02}. It needs a reliable description of its ground state structure, still a challenge after decades of experimental and theoretical efforts \cite{schiffer08,Key09,Mor10}.

In recent years there have been impressive developments in the shell model description of nuclei in the $fp$ shell.  Collective rotational bands and backbending in $^{48}$Cr were described performing shel model calculations in the full $fp$ valence shell \cite{Cau95}. For heavier nuclei, like the neutron-rich Cr, Mn, Fe
and Co isotopes, it has been necessary to include the intruder $0g_{9/2}$ \cite{Lurandi07,Dob08,Kaneko08,Sri09,Sria,Srib,Sun09,Sric}.
The shell model description of the growing experimental evidence for collective behavior of Cr and Fe isotopes with $N \sim$ 40 required the additional inclusion of $1d_{5/2}$ orbital \cite{Nowacki101}. While moving to heavier nuclei it was natural to exclude the $f_{7/2}$ orbital from the valence space, to understand in a shell model context  recent experimental findings in Ga isotopes \cite{Cheal10,Mane11,Diriken10,Cheal10a}, both the proton $f_{7/2}$  and neutron $g_{9/2}$ orbitals were required \cite{Sri12}.

Describing the low energy spectra, electromagnetic and weak transitions in Ge isotopes employing the shell model has many difficulties. The interactions JUN45 \cite{Honma09} and jj44b \cite{brown} are designed to perform shell model calculations in valence space containing  $f_{5/2}, \, p_{3/2}, \, p_{1/2}, \, g_{9/2}$ orbitals. They were fitted to describe the low energy spectra of many isotopes. However, a detailed shell model study of  $^{70,72,74,76}$Ge isotopes by Zamick's group \cite{zamick11} employing both JUN45 and jj44b interactions has shown that the observed $B(E2)$ transition strengths and $Q$ values cannot be properly reproduced. 

Extending Zamick's work, we have enlarged the valence space, including the $\pi f_{7/2}$ orbital to see the effect of proton excitation across Z=28, and extended the calculation up to  $^{82}$Ge, with $N=50$. In the following sections we present a comparison of the low energy spectra of the Ge isotopes with $38\leq N\leq 50$, their quadrupole moments, $B(E2)$ values and occupation numbers, obtained performing shell model calculations with the interactions JUN45 and jj44b in the valence space  $f_{5/2} \, p \,g_{9/2}$ without truncation, with those done in valence $fpg$ space with the effective interaction due to Sorlin {\it et al.} \cite{Sorlin02}, with  $^{48}$Ca as a core and a truncation in the number of excited particles, and compare with the experimental information.

%\newpage

%============================================================================
\section{Details of calculations} 
\vspace{ 12pt}

The present shell model calculations have been carried out in the $f_{5/2} \, p \, g_{9/2}$ and $f \,p \,g_{9/2}$ spaces. 
In the $f_{5/2} \, p \, g_{9/2}$ valence space  the calculations have been performed 
with the interactions JUN45 \cite{Honma09} and jj44b \cite{brown}.
The single-particle energies for the $p_{3/2}$, $f_{5/2}$, $p_{1/2}$ and $g_{9/2}$ orbits 
employed in conjunction with the JUN45 interaction are -9.8280, -8.7087, -7.8388, and -6.2617 MeV respectively. In the case of the  jj44b interaction they are
 -9.6566, -9.2859, -8.2695, and -5.8944 MeV, respectively. The core is $^{56}$Ni, i.e. $N=Z=28$, and the calculations are performed in this valence space without truncation. 

\begin{figure}[h!]
\begin{center}
\includegraphics[scale=0.2]{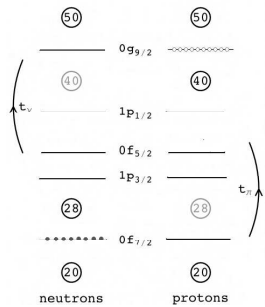}
\caption{Model space $f \,p \,g_{9/2}$ and truncation:
t$_\nu$ neutron excitations from ($p_{3/2}$,$f_{5/2}$,$p_{1/2}$) to $g_{9/2}$, $t_\pi$ proton excitations from $f_{7/2}$ to ($p_{3/2}$,$f_{5/2}$,$p_{1/2}$) orbitals.}
\end{center}
\end{figure}

In the  $f \,p \,g_{9/2}$ valence space,  we use a  $^{48}$Ca core, i.e. only the protons are active in the $f_{7/2}$ orbital, and the interaction $fpg$ reported by Sorlin {\it et al} \cite{Sorlin02}. The single-particle energies are 0.0, 2.0, 4.0, 6.5 and 9.0 MeV for the $f_{7/2}$, $p_{3/2}$, $p_{1/2}$, $f_{5/2}$,  and $g_{9/2}$
orbits, respectively. Since the dimensionality of this valence space is prohibitively large, we have introduced a truncation by allowing $t_\pi$ particle-hole excitations from the $\pi f_{7/2}$ orbital to the
upper  ${fp}$ orbitals ($p_{3/2}$,$f_{5/2}$,$p_{1/2}$) for protons and  $t_\nu$ particle-hole excitations from the upper $fp$ orbitals to the $\nu g_{9/2}$ orbital for neutrons. This is illustrated in Fig. 2. In all the studies reported in this article the maximum allowed value for $t_\pi$ and  $t_\nu$ is four.

%The dimension of
%matrices involved for $f \,p \,g_{9/2}$ space is very large, truncation of the full
%shell model space is necessary. We have allowed neutron: $t_\nu$ jumps from
%($p_{3/2}$, $f_{5/2}$, $p_{1/2}$) to $g_{9/2}$, protons: $t_\pi$ jumps from
%$f_{7/2}$ to ($p_{3/2}$, $f_{5/2}$, $p_{1/2}$) orbital for each nucleus and
%each isotopes. An example of this truncation is shown in Fig. 2. 

 All calculations were carried out at DGCTIC-UNAM computational facility KanBalam and on Tochtli cluster computer at ICN-UNAM, using the shell model
code \textsc{antoine} ~\cite{Antoine}. 
%\newpage
%============================================================================
\section{Energy spectra}
%\vspace{0.2cm}
%\section*{Excitation Energies:}
%\vspace{0.2cm}
%============================================================================

In Figs. 3 to 9 the low energy spectra of the even mass Ge isotopes $^{70, 72, 74, 76, 78, 80, 82}$Ge are presented. In each figure the experimentally determined levels \cite{nndc} are displayed on the left hand side, accompanied by the energy levels obtained through shell model calculations employing the interactions and valence spaces described above, and denoted, from left to right, JUN45, jj44b and fpg.

\subsection{\rm Fig. \ref{Fig3}: $^{70}$Ge}

The first excited $2_1^+$ state is observed at 1039 keV, and predicted by JUN45, jj44b and $fpg$ interactions at 906, 737 and 478 keV, respectively.
The second excited sate is the $0_2^+$, predicted by JUN45 and $fpg$ around it measured energy, but by jj44b at very high in energy.
These first two excited states are reasonably described by the JUN45 interaction. The $fpg$ interaction seems to generate a compressed spectrum for the states with angular momentum 2, 3 and 4, while the energies of the first and second excited $0^+$ states are very close to the experimental ones.

\begin{figure}[h!]
\begin{minipage}{18pc}
\includegraphics[scale=0.40]{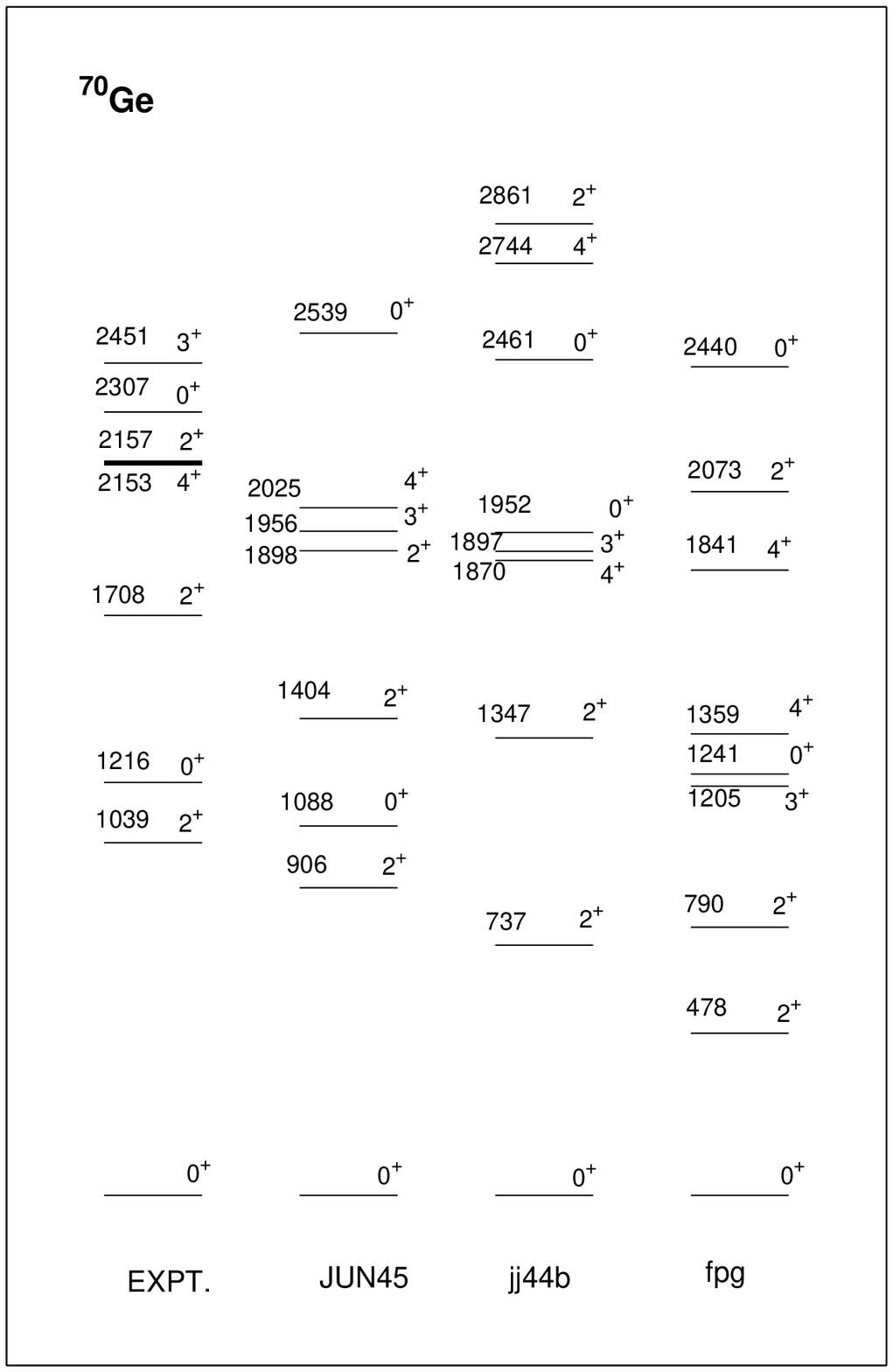}
\caption{\label{Fig3} Calculated and experimental level schemes of $^{70}$Ge, with three different interactions.}
\end{minipage}\hspace{2pc}%
\begin{minipage}{18pc}
\begin{center}
\includegraphics[scale=0.40]{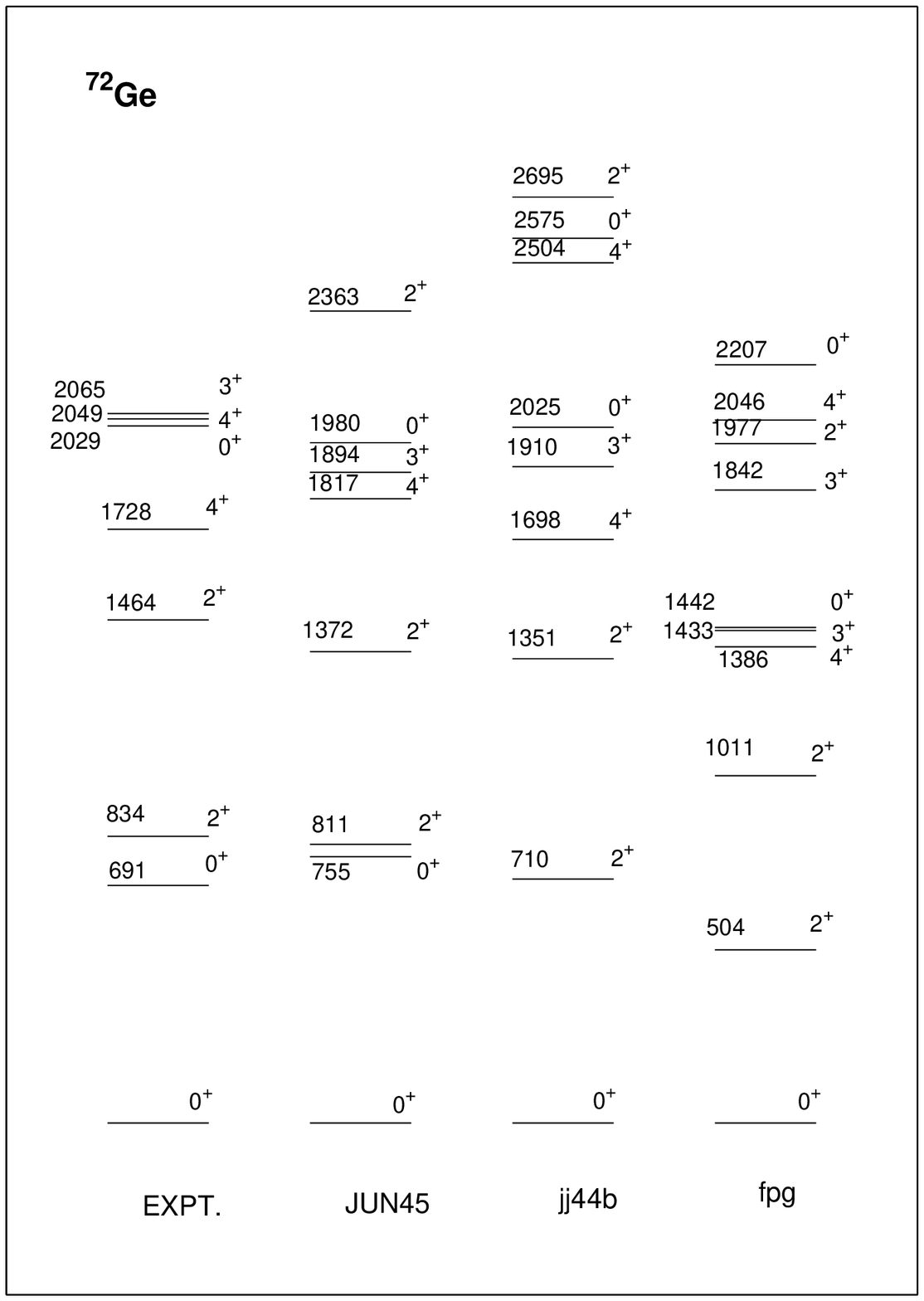}
\caption{\label{Fig4} Calculated and experimental level schemes of $^{72}$Ge, with three different interactions.}
\end{center}
\end{minipage} 
\end{figure}

%============================================================================
\subsection{\rm Fig. \ref{Fig4}: $^{72}$Ge}

As mentioned in the introduction, in $^{72}$Ge the first excited state is the $0_2^+$. Only the JUN45 is able to reproduce this feature. The experimental $2_1^+$ at 834 keV is predicted at 811, 710 and 504 keV, respectively, by JUN45, jj44b and $fpg$ interactions.
Excluding the $0_2^+$ state, the energy spectra obtained with JUN45 and jj44b are similar, while the $fpg$ energies are again compressed for the $2^+, 3^+, 4^+$ states and the energy of the $0_2^+$ state is too high. 

%============================================================================
\subsection{\rm Fig. \ref{Fig5}: $^{74}$Ge}

In this isotope the first two observed excited states have angular momentum 2, followed by the two nearly degenerated states $4_2^+$ and  $0_2^+$.  There is drop in the energy of $2_1^+$, $2_2^+$, and $4_1^+$ from $^{72}$Ge to $^{74}$Ge.
Both JUN45 and jj44b have an acceptable description of the energies of these states. The JUN45 energies are closer to the observed ones, while the jj44b energies have the  $4^+_2$ and  $0^+_2$ in the right order.  The $fpg$ calculations fail in many ways: the energy of the $4_1^+$ state is too low, and places this state below the $2_2^+$, the energy of the $0_2^+$ is very high. 

\begin{figure}[h!]
\begin{minipage}{18pc}
\includegraphics[scale=0.45]{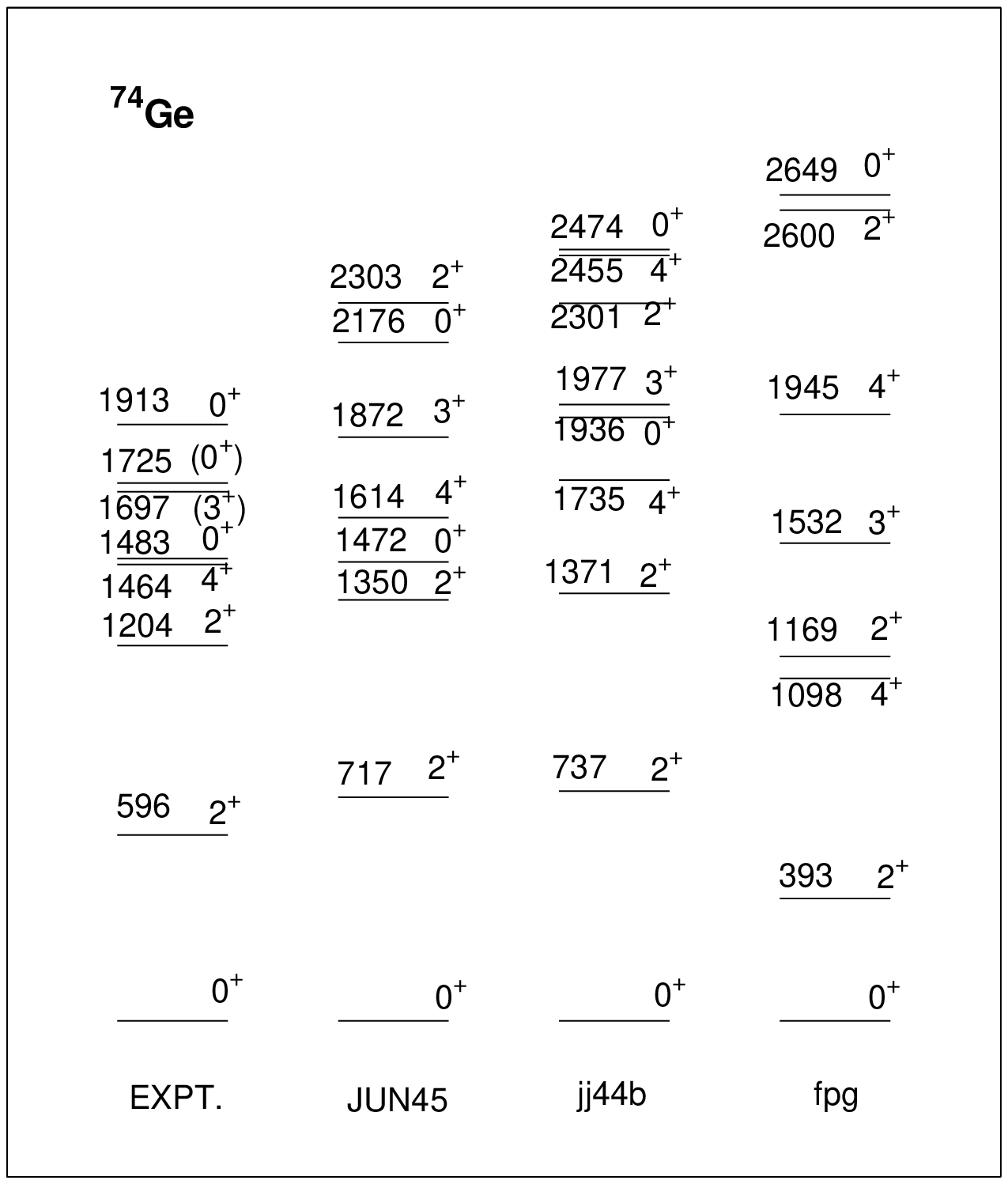}
\caption{\label{Fig5}Calculated and experimental level schemes of $^{74}$Ge, with three different interactions.}
\end{minipage}\hspace{2pc}%
\begin{minipage}{18pc}
\begin{center}
\includegraphics[scale=0.45]{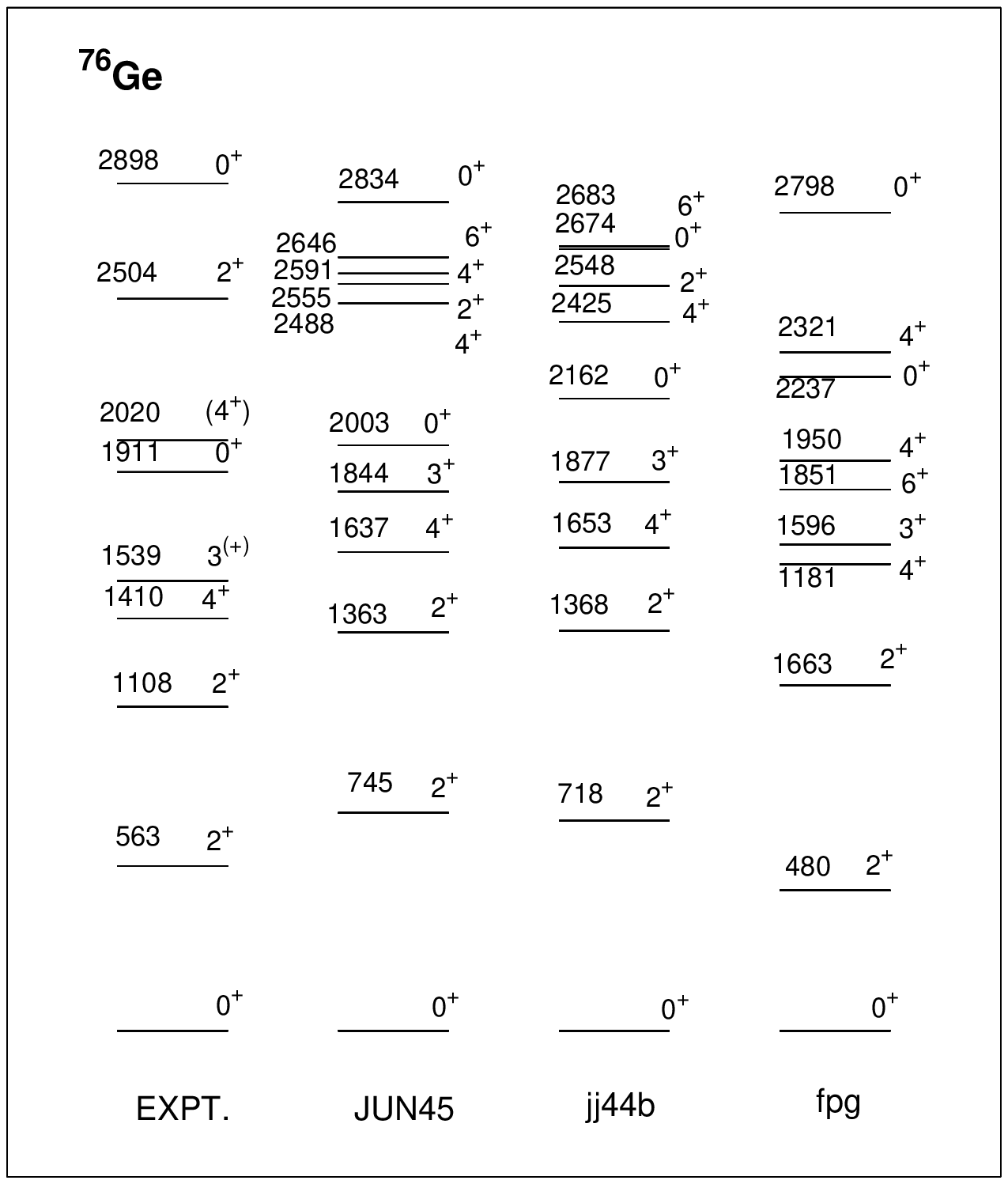}
\vspace{0.65cm}
\caption{\label{Fig6}Calculated and experimental level schemes of $^{76}$Ge, with three different interactions.}
\end{center}
\end{minipage} 
\end{figure}

%============================================================================ 
\subsection{\rm Fig. \ref{Fig6}: $^{76}$Ge}

In this Ge isotope the ordering of the first excited states is  $2_1^+,2_2^+ , 4_1^+, 3_1^+,0_2^+$. The energy spectra obtained using JUN45 and jj44b are very similar and reproduce to the observed ordering, but are displaced to higher energies by 200 keV, with the JUN45 in better agreement with the energy of the $0_2^+$ state. The $fpg$ energies are closer to the observed ones, except for the  $0_2^+$ state which has a higher energy (but closer to the observed one than in for the lighter Ge isotopes), and the $6_1^+$ state, which is predicted below the  $0_2^+$, in a region where there is not reported evidence of its presence. 

%============================================================================
\subsection{\rm Fig. \ref{Fig7}: $^{78}$Ge}

 The ordering of the first excited states in $^{78}$Ge is  $2_1^+, 2_2^+, 0_2^+ , 4_1^+$, with the states $0_2^+$ and  $4_2^+$ very close in energy. The energy spectra obtained with $fpg$ reproduce the observed energies, but displaced to a higher value by about 100 keV. The JUN45 and jj44b results follow the observed trend but the energies are 200 keV too high, and the energy of the $0_2^+$ state is even higher, in particular for JUN45.
 
\begin{figure}[h]
\begin{minipage}{18pc}
\includegraphics[scale=0.45]{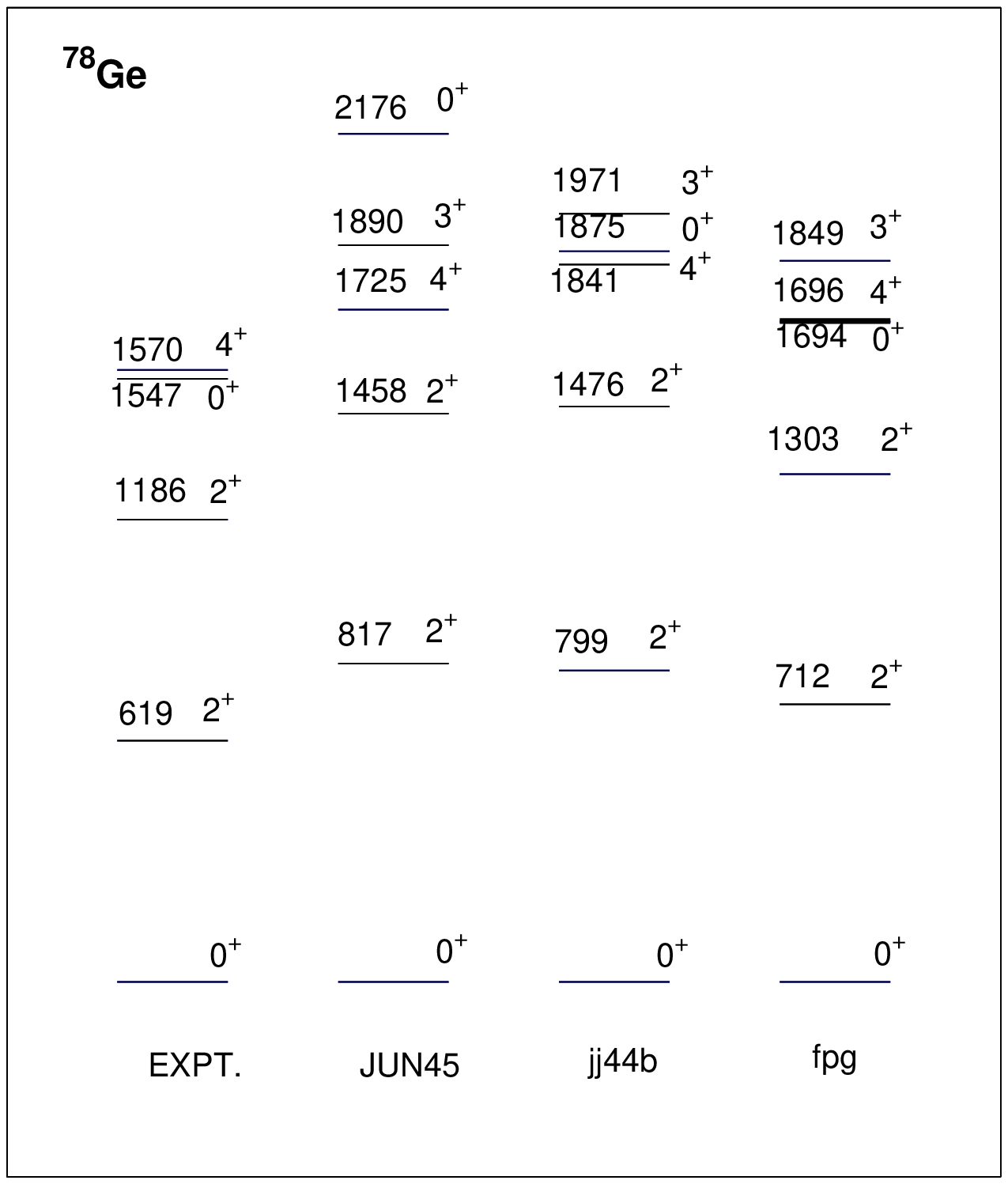}
\caption{\label{Fig7}Calculated and experimental level schemes of $^{78}$Ge, with three different interactions.}
\end{minipage}\hspace{2pc}%
\begin{minipage}{18pc}
\begin{center}
\includegraphics[scale=0.45]{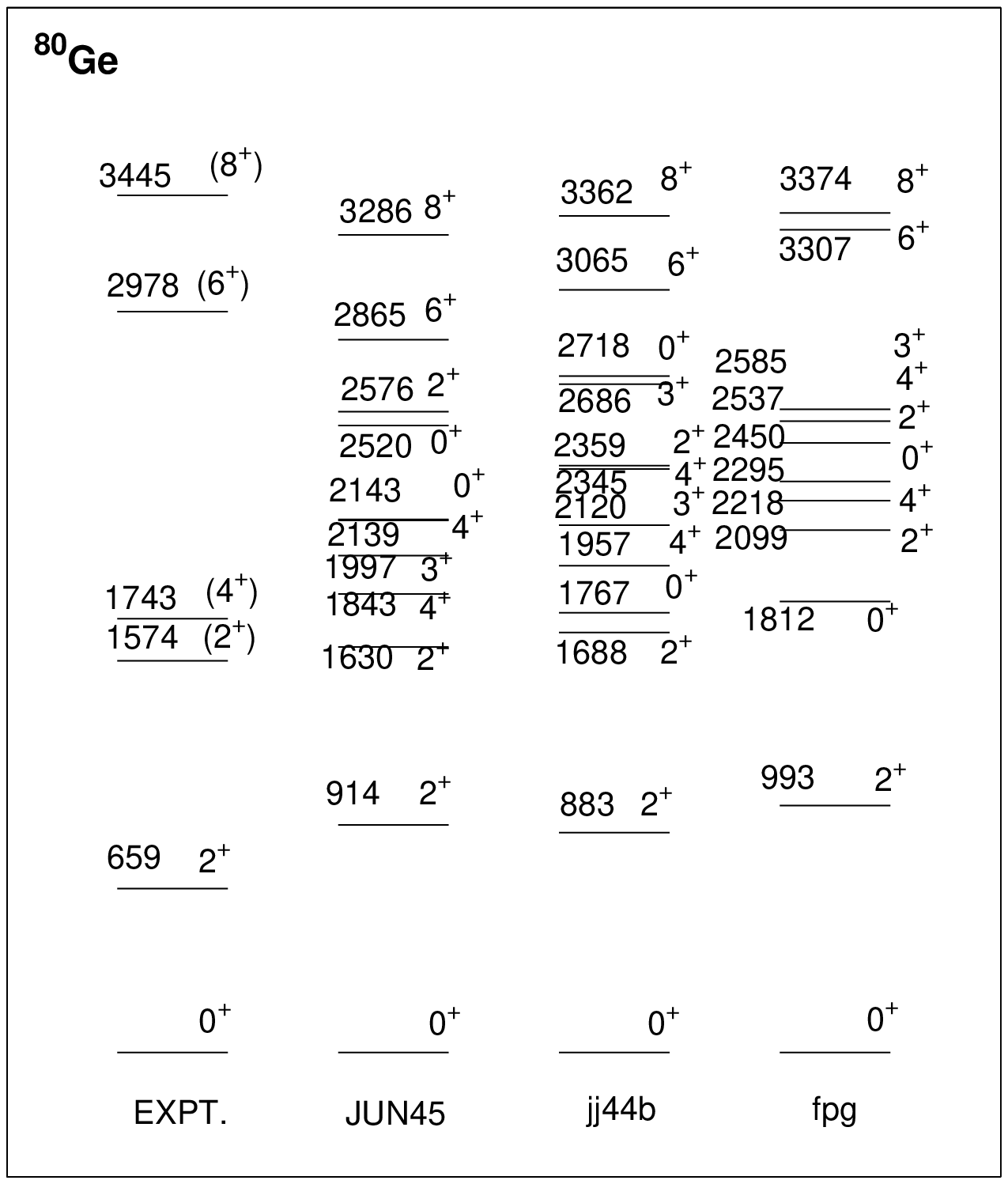}
\caption{\label{Fig8}Calculated and experimental level schemes of $^{80}$Ge, with three different interactions.}
\end{center}
\end{minipage} 
\end{figure}

%============================================================================
\subsection{\rm Fig. \ref{Fig8}: $^{80}$Ge}

For this Ge isotope there are few well determined excited levels.
The three interactions predict the $2_1^+$ state around 300 keV above than experimental data.The $0_2^+$ state has not yet been observed, it is predicted at 2143, 1767, 1812 keV, respectively. 
The experimental information seems to indicate an increase in the excitation energies of the $2_1^+$, $2_2^+$, and $4_1^+$ states, in comparison with $^{78}$Ge. The calculations with three interactions  agree with this trend.
  
\begin{figure}
\begin{center}
\includegraphics[scale=0.45]{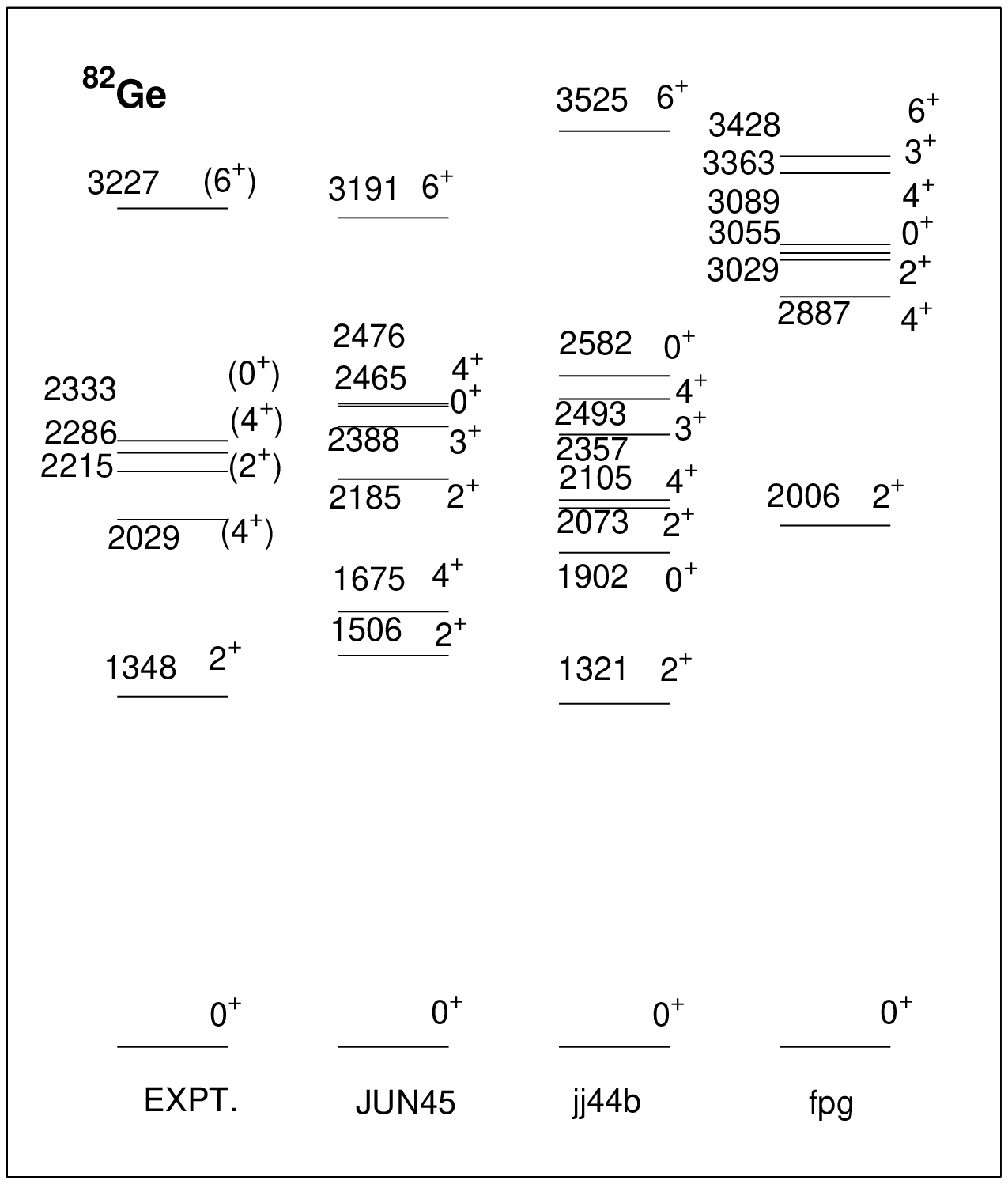}
%\epsfig{file=78Ge.eps, angle=-90, width=0.6\linewidth}
%\resizebox{90mm}{!}{\includegraphics{78Ge_1.eps}}
\caption{\label{Fig9}Calculated and experimental level schemes of $^{82}$Ge, with three different interactions.}
\end{center}
\label{f_82ge}
\end{figure}

%============================================================================
\subsection{\rm Fig. \ref{Fig9}: $^{82}$Ge}

The $^{82}$Ge is semimagic with N=50. The three calculations have no active neutrons, and exhibit serious difficulties to reproduce the main features of the measured levels. The energies of the $2_1^+$,  and $4_1^+$ obtained with JUN45 are too close, the energy of the $0_2^+$ state is too low for jj44b, and the excitations energies are 800 keV too high for fpg.
%============================================================================

\section{ \rm B(E2) Values:}
The electric multipoles  of order $L$ are defined as 
\begin{equation}
         B(el,L)= \frac{1}{2J_{i}+1}\mid(J_f\mid\mid \sum_{i} e_{i}r_{i}^L { Y_{L}}(\theta_{i},\phi_{i})\mid\mid\ J_i)\mid^2,
\end{equation}
where %$L$ is the sum of orbital and intrinsic angular momenta, 
J$_i$ and J$_f$ are the initial and final state spins, respectively.

The $B(E2)$ values is defined as\\

\begin{equation}
        B(E2)= \frac{1}{2J_{i}+1}\mid(J_f\mid\mid \sum_{i} e_{i}r_{i}^2 { Y_{2}}(\theta_{i},\phi_{i})\mid\mid\ J_i)\mid^2. 
\end{equation} 

The experimentally determined $B(E2;2_1^+\rightarrow 0_{g.s.}^+)$ values, associated with the electromagnetic transition strength between the first excited $2^+$ and the ground state, are listed in the second column of Table 1, for the even mass Ge isotopes listed in the first column. The theoretical calculations were performed employing two sets of effective charges:
 e$_{\rm eff}^\pi$ = 1.5 $e$, e$_{\rm eff}^\nu$ = 0.5 $e$ and e$_{\rm eff}^\pi$ = 1.5 $e$, e$_{\rm eff}^\nu$ = 1.1 $e$, displayed in the same column separated by a ``/". 
 
 Except for  $^{82}$Ge, where in the three models the neutrons are not active and do not contribute at all to the transition, leading to small theoretical values, for all the other isotopes the models can reproduce the observed transition intensities. For JUN45 the second set of effective charges works better, jj44b would need some intermediate values for the neutron effective charge, and the large values obtained with $fpg$ are probably related with the contribution of the protons in the $f_{7/2}$ orbital, which could be compensated with smaller effective charges  
 
%The jj44b interaction predicts better $B(E2)$ values for $^{70,72}$Ge in comparison to JUN45 interaction with the first set of effective charges, 
%while for $^{74,76,78,80}$Ge isotopes for jj44b interaction the result with the second set of effective charges is better.
%The JUN45 interaction predicts better result with second set of effective charges.
%The result with $fpg$ interaction is also not much improved. With this interaction the $B(E2)$ values are high in comparison to JUN45 and jj44b interactions. In conclusion, $B(E2)$ results with jj44b %interaction are far better than JUN45 and $fpg$ interactions.
\begin{table}[h]
\caption{The measured \cite{rodal10}  and calculated $B(E2)$ values, given in units of $10^{-3} e^2b^2$.  Two set of effective charges are employed:
 e$_{\rm eff}^\pi$ = 1.5 $e$, e$_{\rm eff}^\nu$= 0.5 $e$ and  e$_{\rm eff}^\pi$ = 1.5 $e$, e$_{\rm eff}^\nu$ = 1.1 $e$, shown separated ``/". }
\bigskip
\begin{center}
%\resizebox{17.5cm}{!}{
\begin{tabular}{|c|c|c|c|c|}
\hline
Nucleus &  Expt. & JUN45 & jj44b & $fpg$\\
\hline
$^{70}$Ge&   36(4)& 24.8/47.5 & 33.7/67.7 &~~ 55.5/99.5~~~~   \\
\hline
$^{72}$Ge& 40(3) & 26.2/50.8& 35.3/71.2 &~~ 44.3/78.1~~~~   \\
        
\hline
$^{74}$Ge &  60(3)& 30.7/59.1 & 36.7/70.6  &~~ 51.5/94.4~~~~   \\
        
\hline
$^{76}$Ge&  46(3) & 31.3/56.9 & 34.8/63.9 &~~ 53.6/93.7~~~~  \\
      
\hline
$^{78}$Ge &  44(3)& 28.5/48.0 & 31.5/53.8  &~~ 48.9/79.8~~~~   \\
        
\hline
$^{80}$Ge & 28(5) & 19.2/30.1 & 22.9/34.8 &~~ 33.3/47.5~~~~   \\
        
\hline
$^{82}$Ge&  25(5)& 8.1/8.1 & 15.9/15.9 &~~ 22.7/22.7~~~~   \\
    
\hline        
\end{tabular}
\end{center}
\end{table} 

%============================================================================
\section{Quadrupole moments:}

The electric quadrupole moment operator is defined as
\begin{equation}
 Q_z= \sum_{i=1}^A {Q_Z(i)} = \sum_{i=1}^A e_i(3z_i^2-r_i^2).
\end{equation} 
The spectroscopic quadrupole moment is defined as
 
\begin{equation}        
  Q_s({J})=\langle  {J},m=  {J} \mid  {Q_2^0} \mid {J},m  = {J}\rangle       
=\sqrt\frac{ {J}(2 {J}-1)}{(2{J}+1)(2{J}+3)({J}+1)}\langle{J}\mid\mid  {Q} \mid\mid {J}\rangle.
  \end{equation} 

The $E2$ operator in Eq. (2) is expressed as a function of the spherical tensor components:
\begin{equation}        
      {Q_2^0}=  Q_z= \sqrt\frac{16\pi}{5}\sum_{i=1}^A e_{i}r_{i}^2Y_{20}(\theta_{i}\phi_{i}).
  \end{equation}

In Table 2  we present the experimental and calculated quadrupole moments, obtained with the three different effective interactions with the two set of effective charges, for the Ge isotopes listed in the first column, and for the first, and in some cases the second, $2^+$ state.
At striking variance with the results obtained for the $B(E2)$ values, the quadrupole moments represent a challenge which surpass the capabilities the three models.
The sign of the quadruple moments is related with the type of deformation: oblate of prolate. It does not change with the rescaling associated with the use of effective charge.
For  $^{72}$Ge the sign of the predicted quadrupole moment is positive for the first $2^+$ state, and negative for the second. The reported experimental signs are the opposite. 
For  $^{74}$Ge only the signs predicted using jj44b coincide with the experimental ones, but the theoretical magnitudes are three to five times smaller. Also for $^{76}$Ge the signs obtained using jj44b are the same give by the experimentalists, and in this case the magnitudes also agree, while the quadrupole moments obtained with JUN45 are very small, and those given by $fpg$ are large with the opposite sign.
The theoretical predictions for  $^{78}$Ge have all negative signs, but differ up to one order of magnitude between the different models.

\begin{table}[h]
\caption{The measured \cite{rodal10}  and calculated quadrupole moments, in $10^{-2} eb$. Two sets of effective charges are employed:
 e$_{\rm eff}^\pi$ = 1.5 $e$, e$_{\rm eff}^\nu$ = 0.5 $e$ and e$_{\rm eff}^\pi$ = 1.5 $e$, e$_{\rm eff}^\nu$ = 1.1 $e$, separated by ``/". }
\begin{center}
\bigskip
%\resizebox{17.5cm}{!}{
\begin{tabular}{|c|c|c|c|c|c|}
\hline
Nucleus & $I$ & Expt. & JUN45 & jj44b & $fpg$ \\
\hline
$^{70}$Ge& $Q(2_1^+)$ &+3(6) or +9(6)& +10.00/+16.94 & +15.13/+24.80  &~~ +29.55/+41.92   \\
\hline
$^{72}$Ge& $Q(2_1^+)$ & -12(8)& +12.83/+21.55& +10.96/+18.61  &~~ +31.28/+44.12 \\
         & $Q(2_2^+)$ & +23(8)& -13.46/-22.27  & -11.31/-19.02 &~~ -32.01/-44.79   \\    
\hline
$^{74}$Ge& $Q(2_1^+)$ &-19(2)& +11.95/+19.86 & -5.89/-6.35 &~~ +39.25/+54.54  \\
         & $Q(2_2^+)$ & +26(6)& -11.46/-18.54  & +5.39/+5.98 &~~ -39.47/-54.56   \\
\hline
$^{76}$Ge & $Q(2_1^+)$ &-14(4)& +1.69/+4.55 & -14.51/-18.63 &~~ +30.04/+40.89  \\
         & $Q(2_2^+)$ & +28(6)&+0.02/-1.94  & +15.52/+20.15 &~~ -29.36/-39.75   \\         
\hline
$^{78}$Ge& $Q(2_1^+)$ & -& -10.90/-13.00 & -15.53/-19.48  &~~ -2.41/-2.21  \\        
\hline          
$^{80}$Ge & $Q(2_1^+)$ & -& -23.45/-28.84 & -23.45/-29.03  &~~ -26.90/-32.11  \\
        
\hline
$^{82}$Ge& $Q(2_1^+)$ & -& -19.56/-19.56 & -18.21/-18.21  &~~ -18.46/-18.46  \\        
\hline 
\end{tabular}
\end{center}
\end{table} 

%============================================================================
\section{Occupation numbers}

\begin{figure}
\begin{center}
\vskip -2.cm
\includegraphics[scale=0.45]{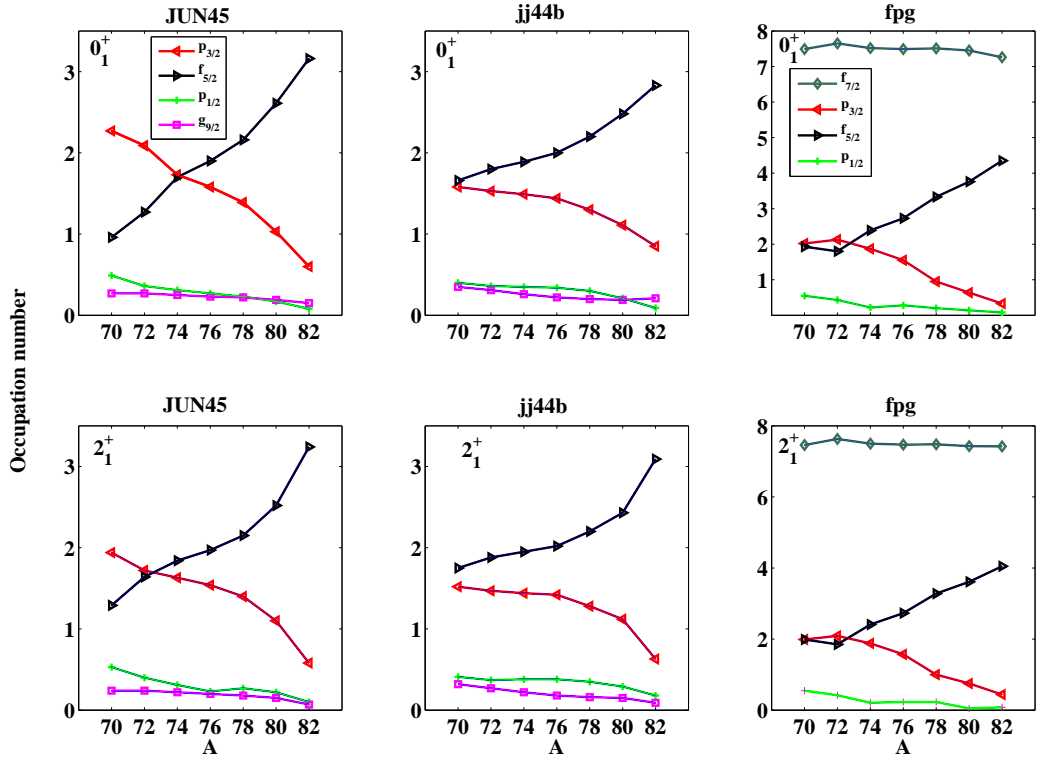}
(a) Proton 
\includegraphics[scale=0.45]{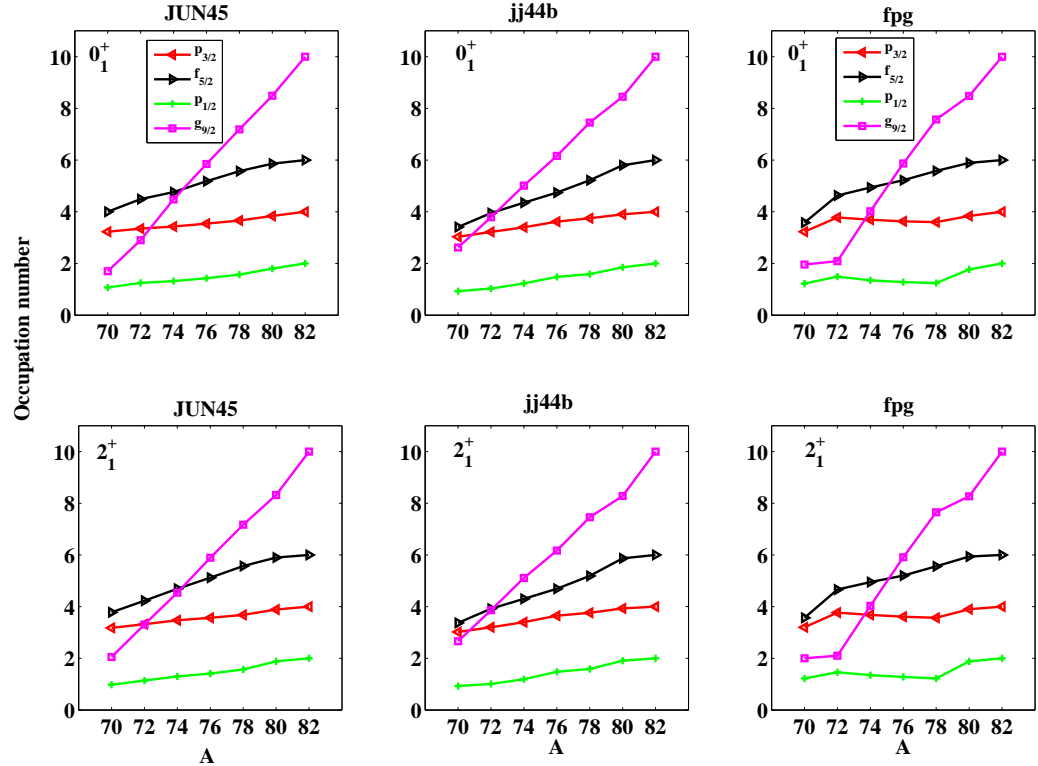}
(b) Neutron 
%\epsfig{file=78Ge.eps, angle=-90, width=0.6\linewidth}
%\resizebox{90mm}{!}{\includegraphics{78Ge_1.eps}}
\caption{\label{Fig10}(Color online) Proton/Neutron occupation numbers of the JUN45 and jj44b  ($p_{3/2}$, $f_{5/2}$, $p_{1/2}$ and $g_{9/2}$ -shell orbits) and $fpg$ ($f_{7/2}$, $p_{3/2}$, $f_{5/2}$, $p_{1/2}$ -shell orbits) interactions-
 for two low-lying states in even-even Ge isotopes. (Upper
panel) $0_1^+$ states;(lower panel) $2_1^+$ states.
}
\end{center}
\label{f_82ge}
\end{figure}

In Fig. \ref{Fig10}, we show the proton (a)  and neutron (b) occupation numbers of the different orbitals for the ground and first excited $2^+$ states of the isotopes $^{70-82}$Ge. 
The general trends are very interesting. The proton occupancies increase smoothly in the $f_{5/2}$ orbital, at the expense of the $p_{3/2}$. It suggests that there is a crossing between these two orbitals due to the interactions with the neutrons. This crossing is predicted to take place at N=38 with jj44b, at N=40 with fpg, and at N=42 employing JUN45.
In the neutron sector increasing the neutron number leads to a nearly linear grow in the occupation of the $g_{9/2}$ orbital, and with a smaller degree of the $f_{5/2}$ occupations.

In Table 3 we have tabulated the occupancies of proton and neutron orbitals for the $^{76,78}$Ge isotopes, for the ground and the first excited $2^+$ states.
Some of these occupancies have been determined experimentally for $^{76}$Ge \cite{schiffer08,Key09}, and these values with their uncertainties are shown in the second row.
In the case of  $^{76}$Ge the proton occupancies predicted with JUN45 and jj44b are in general similar, and both close to those determined experimentally. A clear consequence of opening the $\pi0f_{7/2}$ and closing the  $\pi0g_{9/2}$ orbitals in the $fpg$ calculations is the large occupation of the $\pi0f_{5/2}$, predicting about one proton more than the number measured in this orbital. In the neutron sector
the jj44b occupations fully agree, within the experimental errors, with  the observed ones. The jj44b and JUN45 occupations are quite similar, and both show a slightly larger occupation in the $\nu0f_{5/2}$  at the expense of the $\nu0g_{9/2}$ orbital.
The proton occupations  in $^{78}$Ge exhibit a small displacement from the $\pi1p_{3/2}$ and $\pi1p_{1/2}$ orbitals to the  $\pi0f_{5/2}$ orbital, more noticeable for in the $fpg$ scheme. The neutron occupations are similar for the three interactions.

In both isotopes the occupations calculated for the first excited $2^+$ state are very close to those of the ground state. It is worth to mention that the $\nu0g_{9/2}$ occupations are larger, by nearly to units,  than those expected in the extreme shell model estimation: 4 for  $^{76}$Ge and 6 for $^{78}$Ge. 

\begin{table}[h]
\caption{ Occupation of proton and neutron orbitals for $^{76}$Ge and $^{78}$Ge isotopes in $f_{5/2}pg_{9/2}$ and $fpg_{9/2}$ spaces.}
\begin{center}
\resizebox{16.5cm}{!}{
%\begin{tabular}{|r|c|r||c|r|c|r|c||r|c|r|c|r|}
 \begin{tabular}{|r|c|r||c|r|c|r|c||c|r|c|r|}

\hline
%Interaction &Nucleus &$I$&$\pi0f_{7/2}$&$\pi1p_{3/2}$&$\pi1p_{1/2}$~~&$\pi0f_{5/2}$ ~~~~~~&$\pi0g_{9/2}$&$\nu0f_{7/2}$&$\nu1p_{3/2}$&$\nu1p_{1/2}$~~&$\nu0f_{5/2}$ ~~~~~~&$\nu0g_{9/2}$ \\
Interaction &Nucleus &$I$&$\pi0f_{7/2}$&$\pi1p_{3/2}$&$\pi1p_{1/2}$~~&$\pi0f_{5/2}$ ~~~~~~&$\pi0g_{9/2}$ &$\nu1p_{3/2}$&$\nu1p_{1/2}$~~&$\nu0f_{5/2}$ ~~~~~~&$\nu0g_{9/2}$ \\

\hline
%&&&&\multicolumn{2}{c}{$1p_{1/2}$ +$1p_{3/2}$}   &&\\
Expt.\cite{schiffer08,Key09}~~ &$^{76}$Ge &g.s.& & \multicolumn{2}{c|}
{
%$1p_{3/2}$ +$1p_{1/2}$=
1.75$\pm$0.15}& 2.04$\pm$0.25& 0.23$\pm$0.25%&
& \multicolumn{2}{c|}
{%$1p_{3/2}$ +$1p_{1/2}$=
4.87$\pm$0.20}& 4.56$\pm$0.40&6.48$\pm$0.30\\
\hline
  JUN45~~ &$^{76}$Ge &$0_1^+$& &1.58 &0.27&1.90~~  &0.23 %&
  &3.54 &1.43&5.18~~ &5.85 \\
          &    &$2_1^+$& &1.54 &0.29 &1.97~~ &0.20 %&
          &3.57  &1.41 &5.12~~ &5.89 \\
\hline
 jj44b~~~~ &$^{76}$Ge &$0_1^+$& &1.44 &0.34&2.00~~  &0.22%&
 &3.62  &1.48&4.74~~  &6.16\\
       &     &$2_1^+$& &1.42 &0.38&2.02~~  &0.18 %&
       &3.65  &1.48&4.69~~  &6.17\\
\hline
$fpg$~~~~ &$^{76}$Ge &$0_1^+$ &7.49&1.55  &0.28 &2.73~~ & \-- %0.00 %&8.00
&3.63  &1.28 &5.22~~ &5.87\\
       &  &$2_1^+$ &7.47&1.57 &0.23 &2.73~~ & \-- %0.00 %&8.00
       &3.61  &1.28 &5.20~~ &5.91\\
\hline

\hline
  JUN45~~ &$^{78}$Ge &$0_1^+$& &1.39 &0.23 &2.16~~  &0.22 %&
  &3.66 &1.57 &5.57~~  &7.19 \\
         &  &$2_1^+$& &1.40 &0.27&2.15~~   &0.18 %&
         &3.68 &1.57&5.57~~   &7.17\\
\hline
 jj44b~~~~  &$^{78}$Ge &$0_1^+$& &1.30 &0.30 &2.20~~ &0.20 %&
 &3.75  &1.59 &5.22~~ &7.45\\
        &     &$2_1^+$& &1.28 &0.35&2.20~~ &0.16%&
        &3.76  &1.59 &5.19~~ &7.46 \\
\hline
$fpg$~~~~ &$^{78}$Ge &$0_1^+$& 7.51&0.95&0.20 &3.33~~ & \-- %0.00 %& 8.00
&3.60 &1.24 &5.58~~ &7.57 \\
       &      &$2_1^+$ &7.48&1.00 &0.23&3.28~~ & \-- %0.00 %&8.00
       &3.57 &1.22 &5.56~~ &7.65 \\
\hline\end{tabular}}
\end{center}
\end{table}

%============================================================================
\section{Concluding remarks}

We have presented three sets of shell model calculations for even $^{70-82}$Ge isotopes in the $f_{5/2}pg_{9/2}$ and $fpg_{9/2}$ spaces, employing recently developed interactions. 
The performance of the three calculation schemes is mixed. The JUN45 interaction in the $f_{5/2}pg_{9/2}$ valence space allowed a good description of the observed low energy spectra in  $^{70,72,74}$Ge and $^{80}$Ge, while the $fpg$ interaction in the $fpg_{9/2}$ space worked better for  $^{76,78}$Ge. The calculations for  $^{82}$Ge, having only the proton orbitals active, were very limited in any case.
The $B(E2)$ values were reproduced without much problem in the three schemes, while the quadrupole moments represented a very difficult challenge. The jj44b interaction was able to reproduce the measured quadrupole moments and occupations in  $^{76}$Ge.

One obvious conclusion related to the shell model description of Ge isotopes is the need to include upper orbitals in the neutron sdg shell, following the developments introduced by Lenzi et al \cite{Nowacki101}. The theoretical description of the low energy spectra of isotopes in this mass region would require the inclusion of many orbitals, where we find the known limitation due to the enormous size of the Hilbert space, and the spurious center of mass remotion.

%============================================================================
\section*{Acknowledgements} 
\vspace{0.2cm}
We would like to thank E. Padilla-Rodal  and A. Poves for useful discussions during this work. All calculations were carried out using the computational facilities KanBalam at DGCTIC-UNAM and Tochtli at ICN-UNAM. P. Srivastava is a postdoctoral fellow of DGAPA-UNAM. This work was supported in part by Conacyt, M\'exico, and DGAPA, UNAM project IN103212.

\end{document}